\documentclass[twocolumn,prl,aps,floats,showpacs]{revtex4}
\usepackage{graphicx}
\usepackage{bm}

\begin{document}

\title{Entanglement between static and flying qubits\\ in an Aharonov-Bohm double electrometer}

\author{Henning Schomerus}
\author{John P. Robinson}
\affiliation{Department of Physics, Lancaster University,
Lancaster, LA1 4YB, UK}

\date{January 2007}

\begin{abstract}
We consider the phase-coherent transport of electrons passing
through an Aharonov-Bohm ring while interacting with a tunnel
charge in a double quantum dot (representing a charge qubit) which
couples symmetrically to both arms of the ring. For Aharonov-Bohm
flux $\Phi_{\rm AB}=h/2e$ we find that electrons can only be
transmitted when they flip the charge qubit's pseudospin parity an
odd number of times. The perfect correlations of the dynamics of
the pseudospin and individual electronic transmission and
reflection events can be used to entangle the charge qubit with an
individual passing electron.
\end{abstract}
 \maketitle

\section{Introduction}

A scalable solid-state quantum computer has to rely on a hybrid
architecture which combines static and flying qubits
\cite{nielsen}. Recent works propose the passage of electrons
through
  mesoscopic scatterers for the generation of entanglement
between flying qubits
\cite{loss2000,burkard2000,Costa2001,Oliver2002,bose2002,Saraga2003,beenakkervelsen,buettikersamuelsson,Hu2004},
and charge measurements by an electrometer \cite{pepper}
 have been proposed and realized
to entangle or manipulate static spin and charge qubits
\cite{dqdelectrometers,ruskov,emary,engelloss,trauzettel}. Hence,
it seems natural to exploit mesoscopic  interference and
scattering to entangle flying qubits to static qubits
\cite{jefferson}.

\begin{figure}
\includegraphics[width=\columnwidth]{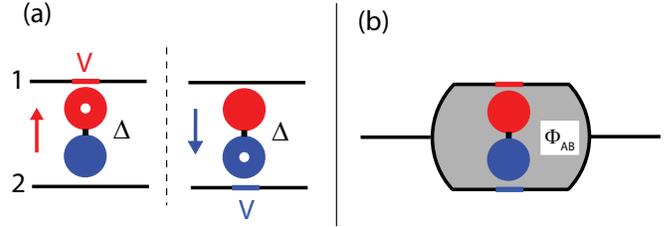}
\caption{(Color online) Left panel: Schematic description of the
double quantum dot (pseudospin $\uparrow$, $\downarrow$ depending
on the occupation of the quantum dots, tunnel splitting $\Delta$),
symmetrically coupled (via the potential $V$) to two quantum
wires. Right panel: The Aharonov-Bohm double electrometer which
results when the quantum wires are pinched together into an
Aharonov-Bohm ring which is pierced
 by a magnetic flux $\Phi_{\rm AB}$.} \label{fig1}
\end{figure}

In this paper we demonstrate that  prompt and perfect entanglement
of a flying and a static charge qubit can be realized when a
double quantum dot occupied by a tunnel charge  is
electrostatically coupled to the symmetric arms of an
Aharonov-Bohm interferometer [an {\em Aharonov-Bohm double
electrometer}, see Fig.\ \ref{fig1}(b)]. For an Aharonov-Bohm flux
$\Phi_{\rm AB}=h/2e$ (half a flux quantum), each electron passing
through the ring signals that the tunnel charge has changed its
quantum state, while each electron which is reflected signals that
the tunnel charge has maintained its state. This can be used to
produce perfect entanglement between the static charge qubit
represented by the tunnel charge in the double dot
\cite{chargequbit}, and the flying qubit represented by the charge
of the conduction electron in the exit leads
\cite{loss2000,burkard2000,Costa2001,Oliver2002,bose2002,Saraga2003,beenakkervelsen,buettikersamuelsson,Hu2004}.
Since this entanglement mechanism does not require any energy fine
tuning (for the implications of energy constraints see Ref.\
\cite{Hu2004}), the entanglement can be produced quickly by the
passage of a single electronic wave packet through the system.

Our proposal draws from both mesoscopic effects mentioned above
--- in essence, it consists of two electrometers both coupling to
the same charge qubit, and  pinched together to form an
Aharonov-Bohm ring, which then represents a mesoscopic scatterer.
Since we require that the coupling of the tunnel charge to the
arms of the ring is symmetric, our proposal falls into the class
of parity meters which have been discussed for the entanglement
and detection of spin qubits \cite{emary,engelloss} and charge
qubits \cite{trauzettel}. In the present paper we are concerned
with charge degrees of freedom only, and the resulting
entanglement should be detectable by current-charge
cross-correlation experiments
\cite{beenakkervelsen,buettikersamuelsson}.

\section{Scattering theory}

In order to describe the properties of the Aharonov-Bohm double
electrometer we start with the scattering region which is depicted
in Fig.\ \ref{fig1}(a). It consists of two  quantum wire segments
labelled 1 and 2, which are symmetrically arranged around the
double quantum dot. The  tunnel charge in the double dot is
described by a pseudospin, associated to states $|\uparrow\rangle$
when the charge is in the upper dot and $|\downarrow\rangle$ when
the charge is in the lower dot. The ground state is given by the
symmetric combination
$|+\rangle=2^{-1/2}(|\uparrow\rangle+|\downarrow\rangle)$, and the
excited state is
$|-\rangle=2^{-1/2}(|\uparrow\rangle-|\downarrow\rangle)$. Both
states are separated by a tunnel splitting energy $\Delta$. The
orbital degree of freedom of the passing electron is described by
 basis states $|1\rangle$ and $|2\rangle$. Via its electrostatic
repulsion potential $V(x)$, the tunnel charge impedes the current
in wire $1$ when it occupies the upper dot, while it impedes the
current in wire $2$ when it occupies the lower dot.

Outside of the range of the  potential $V(x)$ one finds plane-wave
scattering states
\begin{equation}
\psi_{s}=\sum_{\tau,\sigma,\eta=\pm} \psi_{\tau,s}(\eta k_\sigma)
\frac{\exp(i\eta k_\sigma x)}{\sqrt{2k_\sigma}}
(|1\rangle+\tau|2\rangle)\otimes|\sigma\rangle ,
\end{equation}
where $k_\sigma=\frac1\hbar \sqrt{2m(E+\sigma\Delta)}$ for states
of total energy $E$. The symbol $\eta=\pm$ describes the
propagation direction along the wire, $\tau=\pm$ describes the
orbital parity of the passing electron with respect to the wire
segments $1$ and $2$, and $\sigma=\pm$ describes the pseudospin
parity of the state of the double dot. For general scattering
potential $V(x)$, the incoming and outgoing components are then
related by an extended scattering matrix \cite{eplnoat}
\begin{equation}
\widetilde S=\left(%
\begin{array}{cccc}
  \tilde r_{++} & \tilde r_{+-} & \tilde t'_{++} & \tilde t'_{+-} \\
  \tilde r_{-+} & \tilde r_{--} & \tilde t'_{-+} & \tilde t'_{--} \\
  \tilde t_{++} & \tilde t_{+-} & \tilde r'_{++} & \tilde r'_{+-} \\
  \tilde t_{-+} & \tilde t_{--} & \tilde r'_{-+} & \tilde r'_{--} \\
\end{array}%
\right) ,
\end{equation}
which takes the initial and final state of the tunnelling charge
into account:
\begin{equation}
\left(%
\begin{array}{c}
  \psi_{+,l}(-k_+)\\
  \psi_{-,l}(-k_-) \\
  \psi_{+,r}(k_+) \\
  \psi_{-,r}(k_-) \\
\end{array}%
\right) = \widetilde S
\left(%
\begin{array}{c}
 \psi_{+,l}(k_+)\\
  \psi_{-,l}(k_-) \\
  \psi_{+,r}(-k_+) \\
  \psi_{-,r}(-k_-) \\
\end{array}%
\right), \label{eq:stilde1}
\end{equation}
\begin{equation}
\left(%
\begin{array}{c}
  \psi_{-,l}(-k_+)\\
  \psi_{+,l}(-k_-) \\
  \psi_{-,r}(k_+) \\
  \psi_{+,r}(k_-) \\
\end{array}%
\right) = \widetilde S
\left(%
\begin{array}{c}
 \psi_{-,l}(k_+)\\
  \psi_{+,l}(k_-) \\
  \psi_{-,r}(-k_+) \\
  \psi_{+,r}(-k_-) \\
\end{array}%
\right). \label{eq:stilde2}
\end{equation}

Analogously, the complete Aharonov-Bohm double electrometer in
Fig.\ \ref{fig1}(b) is described by an extended scattering matrix
 \begin{equation}
S=\left(%
\begin{array}{cccc}
  r_{++} & r_{+-} & t'_{++} & t'_{+-} \\
  r_{-+} & r_{--} & t'_{-+} & t'_{--} \\
  t_{++} & t_{+-} & r'_{++} & r'_{+-} \\
  t_{-+} & t_{--} & r'_{-+} & r'_{--} \\
\end{array}%
\right) . \label{smat}
\end{equation}
 The
amplitudes $r_{\pm,\pm}$ ($r'_{\pm,\pm}$) are associated to
reflection from the left (right) lead, and the amplitudes
$t_{\pm,\pm}$ ($t'_{\pm,\pm}$) are associated to transmission from
left to right (right to left), while the first (second) subscript
denotes the state of the tunnelling charge after (before) the
passage of the electron through the ring.

The matrix $S$ can be related to the internal scattering matrix
$\tilde S$ by adopting the standard model for an Aharonov-Bohm
ring developed in Ref.\ \cite{imry}. The contacts to the left
($s=L$) and right ($s=R$) lead are characterized by reflection
amplitudes $\alpha_s$ and transmission amplitudes
$\beta_s=\sqrt{1-\alpha_s^2}$ which describe the coupling into the
symmetric orbital parity state $2^{-1/2}(|1\rangle+|2\rangle)$.
The Aharonov-Bohm flux $\Phi_{\rm AB}$ mixes the symmetric and
antisymmetric orbital  parities in the passage from one contact to
the other contact by a mixing angle $\phi=\pi \Phi_{\rm AB} e/h$.
The lengths of the ballistic regions between the scattering region
and the left and right contact are denoted by $d_L$ and $d_R$,
respectively.

The total scattering matrix is then of the form
\begin{eqnarray}
&&S=-\alpha + \beta(A^\dagger-\alpha -B^\dagger
[A^\dagger+1]^{-1}B^\dagger )^{-1}\beta , \nonumber
\\
&&A=\gamma\left(%
\begin{array}{cccc}
  \tilde r_{++} & 0 & c\tilde t'_{++} & -is\tilde t'_{+-} \\
  0 & \tilde r_{--} & -is\tilde t'_{-+} & c\tilde t'_{--} \\
  c\tilde t_{++} & is\tilde t_{+-} & \tilde r'_{++} & 0 \\
  is\tilde t_{-+} &  c\tilde t_{--} & 0 & \tilde r'_{--} \\
\end{array}%
\right) \gamma , \nonumber
\\
&&B= \gamma
\left(%
\begin{array}{cccc}
  0 & \tilde r_{+-} & -is \tilde t'_{++} & c \tilde t'_{+-} \\
  \tilde r_{-+} & 0 & c \tilde t'_{-+} & -is \tilde t'_{--} \\
  is \tilde t_{++} & c \tilde t_{+-} & 0 &  \tilde r'_{+-} \\
  c \tilde t_{-+} & is \tilde t_{--} & \tilde r'_{-+} & 0 \\
\end{array}%
\right) \gamma , \label{smatresult}
\end{eqnarray}
$\alpha={\rm diag}(\alpha_L,\alpha_L,\alpha_R,\alpha_R)$,
$\beta={\rm diag}(\beta_L,\beta_L,\beta_R,\beta_R)$, $\gamma={\rm
diag}\,(e^{ik_+d_L},e^{ik_-d_L},e^{ik_+d_R},e^{ik_-d_R})$,
$c=\cos\phi$, $s=\sin\phi$.

\section{Degree of entanglement}
\subsection{Stationary concurrence}

The scattering amplitudes of the extended scattering matrix
(\ref{smat}) can now be used to assess how the tunnel charge on
the double dot becomes entangled with the itinerant electron
during its passage through the Aharonov-Bohm double electrometer.
In particular, they describe by which lead the passing electron
exits and how this is correlated to the final state of the double
dot. We assume that the double dot is initially in the symmetric
state $|+\rangle$ and hence uncorrelated to an arriving electron
that enters the ring from the left lead. The degree of
entanglement between the final state of the double dot and the
exit lead of the electron can then be quantified by the
concurrence \cite{concurrence}
\begin{equation}
{\cal C}=2|r_{++}t_{-+}-r_{-+}t_{++}|, \label{eq:concurrence}
\end{equation}
which provides a monotonous measure of entanglement (${\cal C}=0$
for unentangled states and ${\cal C}=1$ for maximal entanglement).

For the  case of a vanishing flux $\Phi_{\rm AB}=0$, the mixing
angle is $\phi=0$.  It then follows from Eq.\ (\ref{smatresult})
that the scattering matrix is of the form
\begin{equation}
S=\left(%
\begin{array}{cccc}
  r_{++} & 0 & t'_{++} & 0 \\
  0 & r_{--} & 0 & t'_{--} \\
  t_{++} & 0 & r'_{++} & 0 \\
  0 & t_{--} & 0 & r'_{--} \\
\end{array}%
\right) . \label{smatphi0}
\end{equation}
In this case the electron can only leave the system when the
pseudospin of the scatterer has flipped an even number of times,
hence the final state of the scatterer is identical to its initial
state. There is no entanglement, and consequently the concurrence
(\ref{eq:concurrence}) vanishes, ${\cal C}=0$.

For $\Phi_{\rm AB}=h/2e$, the  mixing angle is $\phi=\pi/2$, and
the scattering matrix is of the form
\begin{equation}
S=\left(%
\begin{array}{cccc}
  r_{++} & 0 & 0 & t'_{+-} \\
  0 & r_{--} & t'_{-+} & 0 \\
  0 & t_{+-} & r'_{++} & 0 \\
  t_{-+} & 0& 0  & r'_{--} \\
\end{array}%
\right)
.
\label{smatphipi}
\end{equation}
 Now the electron becomes entangled with the
double dot: the electron is reflected when the double dot is
finally in its symmetric state (the pseudospin of the double dot
has then flipped an even number of times), while the electron is
transmitted when the double dot is finally in its antisymmetric
state (it then has flipped an odd number of times).
 The concurrence ${\cal C}=2|r_{++}t_{-+}|$
signals perfect entanglement, ${\cal C}=1$, when the probabilities
of reflection and transmission are identical,
$|r_{++}|^2=|t_{-+}|^2=1/2$. The condition for perfect
entanglement corresponds to the case of maximal shot noise
\cite{beenakkervelsen,buettikersamuelsson}.

The entanglement in the Aharonov-Bohm double electrometer results
because the total parity $\sigma\tau$ is conserved in the
scattering from the double quantum dot. This conservation law
yields the separate Eqs. (\ref{eq:stilde1},\ref{eq:stilde2}),
which only couple wave components of the same total parity. Since
the scattering matrix $\widetilde S$ for positive and negative
total parity is identical, this  realizes a parity meter which
entangles the orbital parity of the passing electron to the
pseudospin parity of the double dot. The role of the Aharonov-Bohm
ring with flux $\Phi=h/2e$ is to convert the orbital parity into
charge separation in the exit leads. When the electron enters the
ring at the left contact, this prepares it locally in the
orbitally symmetric state. Neglecting the scattering from the
double dot, the electron cannot leave at the right contact since
it will end up there in the antisymmetric orbital state.
Transmission is therefore only possible if the orbital parity is
flipped by interaction with the tunnel charge. But since the total
parity is preserved, this scattering event is necessarily
accompanied by a parity change of the double dot. Consequently,
the final state of the tunnel charge is perfectly correlated to
the lead by which the passing electron leaves the Aharonov-Bohm
double electrometer.

\subsection{Time-dependent concurrence}

The energy-dependent scattering matrix (\ref{smat}) and the
expression (\ref{eq:concurrence}) for the concurrence describe the
stationary transport through the system.  A pressing issue for
many proposals involving flying qubits is that they require an
energy fine tuning which entails a degrading of the entanglement
in the time domain \cite{Hu2004}. As we will now show, the
entanglement can in fact be {\em increased} above the value of the
stationary concurrence when a single electronic wave packet is
passed through the Aharonov-Bohm electrometer with $\Phi_{\rm
AB}=h/2e$.

In this non-stationary situation, the entanglement is quantified
by the time-dependent concurrence
\begin{equation}
{\cal C}(t)=2\Big
|\langle\psi_+|\psi_+\rangle\langle\psi_-|\psi_-\rangle-
|\langle\psi_+|\psi_-\rangle|^2\Big|^{1/2},
\end{equation}
where $\psi_+$ and $\psi_-$ are the orbital wavefunctions obtained
by projection onto the symmetric and antisymmetric state of the
double dot, respectively. The potential for entanglement
enhancement
 follows from the {\em lower} bound
\begin{equation}
{\cal C}(t)>{\cal C}_{\rm
min}(t)=2\sqrt{P_L(t)P_R(t)-\frac{1}{4}P_0(t)^2}
 \end{equation}
 of the time-dependent concurrence, which can be  derived directly from the  form
(\ref{smatphipi}) of the scattering matrix when one assumes that
the wave packet has completely entered the ring.
 Here $P_{L}(t)$ is the weight of the reflected wave packet,
 $P_{R}(t)$ is the weight of the transmitted wave packet, and
 $P_0(t)=1-P_L(t)-P_R(t)$ is the weight of the wave packet in the
 ring.
For large times $P_0(t=\infty)=0$, and the weights
$P_L(t=\infty)=R$ and $P_R(t=\infty)=T$ give the reflection and
transmission
 probabilities of the wave packet, which can be obtained by an energy
 average over the wave packet components.
For the case that this energy average yields equal reflection and
transmission probabilities $R=T=1/2$, the final value of the
concurrence after passage of the electron is $C(t=\infty)=1$. In
this case, maximal entanglement results during the passage of the
electron through the system -- independent of the precise
dependence of the stationary concurrence in the energetic range of
the wave packet. This entanglement enhancement is only possible
because the scattering matrix $S$ retains its sparse structure
(\ref{smatphipi}) for all energies.

To give a specific example, we consider the case that the
tunnelling charge induces a localized scattering potential
$V=\frac{\hbar^2}{2m}g\delta(x)$ in the quantum wire. This problem
can be solved exactly for arbitrary values of the tunnel splitting
$\Delta$, scattering strength $g$, and ring parameters $d_s$ and
$\alpha_s$, using the technique of Ref.\ \cite{eplnoat}. Here we
concentrate on the case of a vanishing tunnel splitting
$\Delta=0$, strong scattering $g\to\infty$, transparent contacts
with $\alpha_R=\alpha_L=0$, and equal distance $d_L=d_R\equiv d$
of the double dot to the contacts. The extended scattering matrix
is then of the form
\begin{equation}
S=-e^{3ikd}\left(%
\begin{array}{cccc}
  \tilde c & 0 & i \tilde s c   & -\tilde s s  \\
 0 & \tilde c& -\tilde s s  & i \tilde s c    \\
  i \tilde s  c   & \tilde s s  &\tilde c & 0 \\
  \tilde s s   & i \tilde s c    & 0 & \tilde c \\
\end{array}%
\right), \label{eq:smatdelta}
\end{equation}
where $\tilde c= \cos kd$ and $\tilde s= \sin kd$.

At fixed energy, the stationary concurrence (\ref{eq:concurrence})
is given by ${\cal C}=|\sin\phi \sin 2kd|$, and the maximal
stationary concurrence ${\cal C}=|\sin 2kd|$ is attained at
$\phi=\pi/2$, corresponding to $\Phi_{\rm AB}=h/2e$. The factor
$|\sin 2kd|$ arises from the energy dependence of the reflection
probability $R=\cos^2 kd$.  For a narrow wave packet, the
reflection and transmission probabilities average to $1/2$, and
the time-dependent concurrence  approaches the maximal value
${\cal C}(t=\infty)=1$ for large times.

\begin{figure}
\includegraphics[width=\columnwidth]{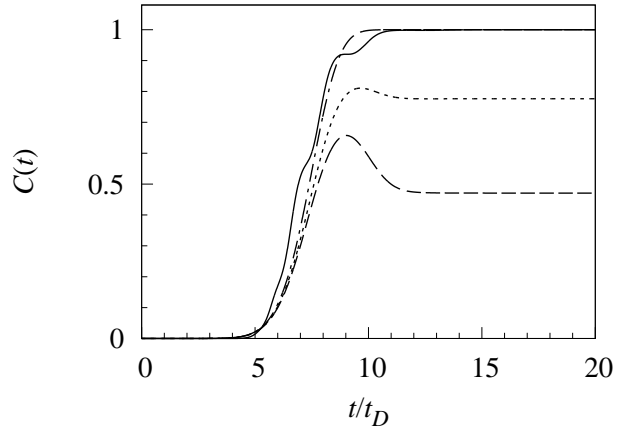}
\caption{Time-dependent concurrence when an electronic wave packet
passes through the ring with Aharonov-Bohm flux $\Phi_{AB}=h/2e$,
while the double quantum dot is initially in its symmetric state.
The scattering potential is $V=\frac{\hbar^2}{2m}g\delta(x)$ with
$g\to\infty$. The contacts to the ring are  transparent,
$\alpha_R=\alpha_L=0$. The distance $d_L=d_R\equiv d$ of the
double dot to the contacts is $d=125\,a$ , where $a$ is the lattice
constant. Time is measured in units of the average time of flight
$t_D=2d/\langle v\rangle$ of the wave packet between the two
contacts. The center of the initial wave packet in the lead is at
a distance of $2000\,a$ to the ring. The initial widths $w$ and the
average wave numbers $\langle k\rangle$ are: $w=0.2 d$, $\langle
k\rangle d=4\pi$ (solid curve), $w=2 d$, $ \langle k\rangle d=
4\pi$ (dashed curve), $w=2 d$, $\langle k\rangle d=4.125\pi$
(dotted curve), $w=2 d$, $\langle k\rangle d=4.25\pi$
(dashed-dotted curve).} \label{fig2}
\end{figure}

In order to assess the time scale on which the entangled state is
formed we have performed
 numerical simulations which are presented
in Fig.\ \ref{fig2}. The wave packet was propagated by a
second-order Crank-Nicholson scheme, and the Aharonov-Bohm ring
and the leads were formed by tight-binding chains at wavelengths
much larger than the lattice constant. For a narrow wave packet,
the simulations confirm that the final state is perfectly
entangled. For broader wave packets, the final entanglement is not
necessarily perfect and depends on the average wave number, but is
always attained on a time scale comparable to the propagation time
of the wave packet between the contacts in the ring.

\begin{figure}
\includegraphics[width=\columnwidth]{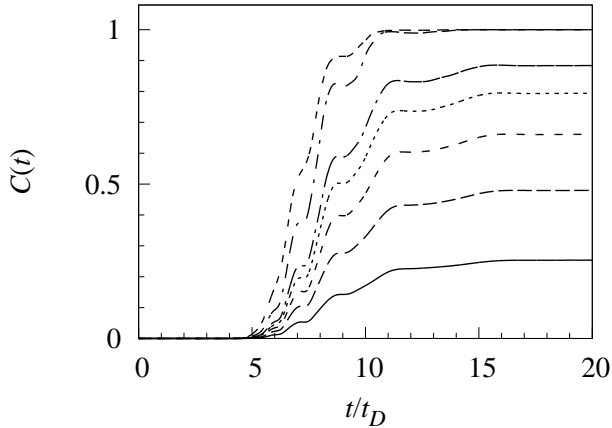}
\caption{Time-dependent concurrence for the narrow wave packet
with $w=0.2 d$ and $\langle k\rangle d=4\pi$ described in Fig.\
\ref{fig2}, but now calculated for finite values of the coupling
strength $g=0.001$, $0.002$, $0.003$, $0.004$, $0.005$, $0.01$,
$0.1$ (bottom to top).} \label{fig3}
\end{figure}

\begin{figure}
\includegraphics[width=\columnwidth]{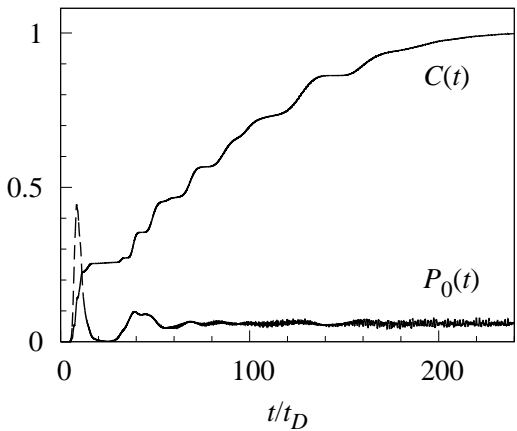}
\caption{Solid line: Time-dependent concurrence for the narrow
wave packet with $w=0.2 d$ and $\langle k\rangle d=4\pi$ described
in Fig.\ \ref{fig2}, but for $g=0.001$ and a set-up in which the
wires are terminated by hard walls at a distance $4000\,a$ to
either side of the ring. The concurrence is enhanced due to the
multiple passages of the wave packet through the ring. The dashed
line shows the probability $P_0(t)$ of the propagating electron to
reside in the ring, which is revived in the multiple passages
through the ring.} \label{fig4}
\end{figure}

In Fig.\ \ref{fig3} we assess how the  degree of entanglement for
a narrow wave packet depends on the scattering strength $g$.
Perfect entanglement is obtained for $g\gtrsim 0.1$, hence,
already for rather weak coupling. For smaller $g$, the
entanglement produced by a single passage of the wave packet is
reduced. As required, the entanglement vanishes altogether  in the
limit $g\to 0$. For small but finite $g$, one should expect that
the entanglement can be enhanced by multiple passage of the wave
packet through the ring, which can be achieved by isolating the
system for a finite duration from the external electrodes (e.g.,
by pinching off the external wires via the voltage on some split
gates). This expectation is confirmed in Fig.\ \ref{fig4}, which
shows the time-dependent concurrence for a narrow wave packet
which passes through the ring and is reflected at hard-wall
boundaries in the external wires, at a distance of $4000\,a$ to
either side of the ring.

\subsection{Sensitivity to decoherence and asymmetries}

\begin{figure}
\includegraphics[width=\columnwidth]{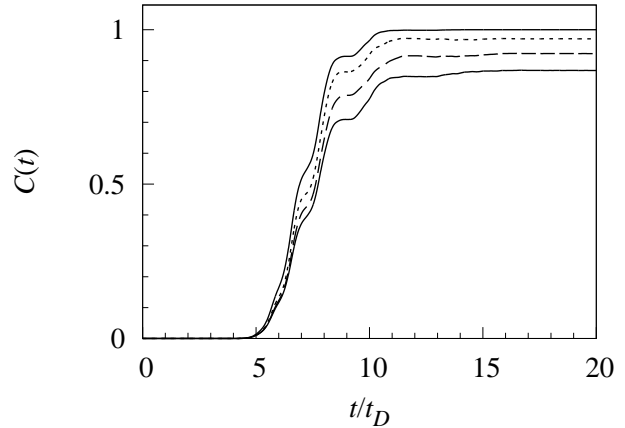}
\caption{Time-dependent concurrence for the narrow wave packet
with $w=0.2 d$ and $\langle k\rangle d=4\pi$ described in Fig.\
\ref{fig2}, but for asymmetric coupling to the double dot. In one
arm, the coupling strength is $g=0.1$ while in the other arm it is
reduced by  a factor $\alpha=1$, $0.1$, $0.05$, and $0$ (top to
bottom).} \label{fig5}
\end{figure}

The time-dependent scattering analysis of the entanglement
mechanism reveals an important and rather unique feature of the
proposed device: As the entanglement can be generated in a single
quasi-instantaneous scattering event, the proposed mechanism is
rather robust against decoherence from time-dependent fluctuations
of the environment (decoherence, however, will always become
important in the subsequent dynamics of the system
\cite{nielsen}). The main source of entanglement degradation hence
comes from imperfections in the fabrication of the device itself,
and here especially from imperfections  which break the parity
symmetry of the set-up. Arguably the most critical part of the
set-up is the requirement of symmetric coupling to the double dot,
as other asymmetries (in the arm length and contacts to the
external wires) are just of the same character as in the
conventional Aharanov-Bohm effect and can be partially
compensated, e.g., via off-setting the magnetic flux. The
sensitivity to asymmetries in this coupling is explored in Fig.\
\ref{fig5}, where the coupling strength to one arm is fixed to
$g=0.1$ while in the arm the strength is reduced by a factor
$\alpha$. Astonishingly, a rather large degree of entanglement
remains even for much reduced $\alpha$, which indicates that the
proposed mechanism is rather more robust to asymmetries than it
could have been expected.

\section{Discussion and conclusions}

In conclusion, we have demonstrated that a tunnel charge on a
double quantum dot can be  entangled to an individual electron
which passes through an Aharonov-Bohm ring with flux $\Phi_{\rm
AB}=h/2e$. Ideal operation requires symmetric electrostatic
coupling of the tunnel charge to the two arms of the ring. The
entangled state is produced quickly, on time scales comparable to
the passage of the electron through the ring. Since the
entanglement in principle is generated in a single
quasi-instantaneous scattering event, the proposed mechanism is
robust against decoherence.

The mesoscopic components for the proposed entanglement circuit --
double quantum dots, ring geometries of quantum wires, and charge
electrometers based on the electrostatic coupling of quantum dots
to quantum wires -- have been realized and combined in numerous
experiments over the past decade, and recently the dynamics of
static qubits in double quantum dots has been monitored
successfully \cite{chargequbit,dqdelectrometers}. An initial
experiment would target the finite conductance of the mesoscopic
ring at half a flux quantum, $\Phi_{\rm AB}=h/2e$, where the
conventional Aharonov-Bohm effect would yield total destructive
interference and hence no current. The shot noise provides an
indirect measurement of the underlying correlated dynamics of the
tunnel charge and the mobile electron. A direct experimental
investigation of the ensuing entanglement requires to measure the
correlations between the individual transmission and reflection
events and the pseudospin dynamics of the double quantum dot. The
cross-correlation measurements could be carried out with a
stationary stream of electrons, at a small bias $V<t_D/e\hbar$
(hence an attempt frequency less than the typical dwell time $t_D$
of an electron in the ring). As in other proposals involving
flying qubits, the incoming stream of electrons can be further
diluted by a tunnel barrier in the lead from the electronic
source. In a more sophisticated setting, a single charge would be
driven into the system via an electronic turnstile
\cite{turnstile}. It also may be desirable to lead the reflected
or transmitted charges into dedicated channels via electronic beam
splitters or chiral edge states \cite{Ji}.

We thank C. W. J. Beenakker and T. Brandes for helpful discussions.
This work was supported by the European Commission, Marie Curie
Excellence Grant MEXT-CT-2005-023778 (Nanoelectrophotonics).

\end{document}